\documentclass[article]{raa}
\usepackage{graphicx,times,amsmath,amssymb,natbib}             
\usepackage{natbib}
\bibpunct{(}{)}{;}{a}{}{,}
\usepackage{color}
\usepackage{subfigure}
\usepackage{upgreek}

\usepackage[a4paper=true,dvipdfm=true,pagebackref=true]{hyperref}
\hypersetup{pdftitle = The title of my PDF, pdfauthor = My name, pdfsubject= The subject, pdfkeywords = keyword1 keyword2 keyword3} 
\hypersetup{colorlinks = true, linkcolor = green, anchorcolor = red, citecolor = blue, filecolor = red, pagecolor = red, urlcolor = red}
\begin{document}
\title{Detection of the thermonuclear X-ray bursts and dips from the \textbf{X-ray binary} 4U 1323-62 with \textit{AstroSat/LAXPC}.}
 \volnopage{Vol.0 (200x) No.0,000--000}    
       \setcounter{page}{1}          

\author{Yashpal Bhulla\inst{1}  \and Jayashree Roy\inst{2} \and S.N.A Jaaffrey\inst{1} } 
\institute{ Pacific Academy of Higher Education and Research University, Udaipur-313003, India (yash.pkn@gmail.com) \\
\and Inter-University Center for Astronomy and Astrophysics, Pune-411007, India \\
}

\abstract{
Using data from the Large Area X-ray Proportional Counter (LAXPC) on the \textit{AstroSat} satellite, we observed the Type-1 thermonuclear X-ray bursts from the  Low Mass X-ray Binary Neutron Star 4U 1323-62. The observations of 4U 1323-62 were carried out during the performance verification phase of the  \textit{AstroSat} satellite showed six thermonuclear X-ray bursts in a total effective exposure of $\sim$ 49.5 ks for about two consecutive days. Recurrence time of the detected thermonuclear bursts is in accordance with the orbital period of the source $\sim$ 9400 seconds. Moreover, the light curve of 4U 1323-62 revealed the presence of two dips. We presented the results from time-resolved spectroscopy performed during all of the six X-ray bursts and also report the detection of a  known low frequency Quasi-periodic oscillation (LFQPO) at $\sim$ 1 Hz, from the source. However, any evidence of kilohertz QPO was not found. We have shown the burst profile at different energy ranges. Assuming a distance of 10 kpc, we observed a mean flux $\sim$ 1.8 $\times$ $10^{-9}$ erg $cm^2$ $s^{-1}$. The radius of the blackbody is found to be highly consistent with the blackbody temperature and the blackbody flux of the bursts.
}
\authorrunning{Bhulla et al.}
\titlerunning{thermonuclear X-ray bursts : 4U 1323-62}

\maketitle
\section{Introduction}
 Low mass X-ray binaries (LMXBs) are the system in which the compact object neutron star (NS) or a black hole (BH) accretes matter from the companion star via the Roche lobe overflow \citep{1989hasi,1998urpin,2006klis, 2011reig,2014belloni,motta2016,Chauhan_2017,2019bhulla}. In LMXBs number of periodic and quasi-periodic oscillations (QPOs) have been observed \citep{1999Mendez,2000van,2000mendez,2001mendez,2005barret}. These QPOs are attributed to the instabilities in the accretion disk. \citep{2005homan,2011ingram,2015scaringi}. LMXBs span a broad range of luminosities from $\sim$ $10^{36}$ to a few $10^{38}$ erg $s^{-1}$ \citep{2014churuch,2016wangj}. The weakly magnetized and luminous NS LMXBs are classified into Z and atoll sources based on their color-color diagram (CCD) and hardness intensity diagram (HID) \citep{1989hasi,2018mond}.

It has been observed in some of the neutron star sources that the X-ray luminosity increases by a factor of $\sim$ 10 in about a few seconds, approaching Eddington luminosity $L_{EDD}$ = $\sim$ $10^{38}$ ergs $s^{-1}$ band then fades with a timescale of $\sim$ 10  seconds. This phenomenon is known as thermonuclear type-I X-ray bursts \citep{2006stroh,2008galloway}. The first type-I X-ray burst was discovered in the 1970s. The Multi-INstrument Burst ARchive (MINBAR)\footnote{http://burst.sci.monash.edu/minbar/} is a database containing the analysis of more than 7000 type-I thermonuclear X-ray bursts from 84 burst sources, that are observed by the different X-rays observatories \textit{BeppoSAX}, \textit{Rossi X-ray Timing Explorer (RXTE)}, \textit{INTEGRAL}. Type-I X-ray bursts are flashes due to the unstable thermonuclear burning of accreted and accumulated material on a neutron star \citep{1976grindlay,1976belian,1998bildsten,2006stroh,2017galloway,Beri_2018}. This ignition condition is caused due to hydrogen and/or helium enrich component supplied by a companion star to proceed with the burning \citep{1975hansen,1976wooslay,1977maraschi,1978lamb, 1993lewin,2006stroh}. Bursts generally appear as a short transient event where the X-ray intensity increases rapidly on a time scale of few seconds, and decays in an exponential trend back to the pre-burst level. The decay time is always more than the rise time, and the duration of thermonuclear burst range from a few seconds to half an hour. The time interval of burst-to-burst events is typically in the order of hours to days. The time interval is supposed to accumulate enough fuel to generate another burst. After the X-ray eruptions, luminosity recedes to its pre-burst stage \citep{1984lewin,1984tawara}.

Thermonuclear X-ray bursts have been studied with the \textit{Rossi X-ray Timing Explorer/Proportional Counter Array} (\textit{RXTE/PCA}) \citep{2003gallow,2004gallo}. The shortest recurrence time  for neutron star 4U 1636-536 is $\sim$  5.4 minutes reported by \cite{2009lina} by using  \textit{RXTE/PCA} observations. \cite{2010keek} observed the shortest recurrence time from 4U 1705-44 to be $\sim$ 3.8 minutes. MINBAR locate the 61 X-ray burst sources by \textit{RXTE/PCA}, 56  burst sources by \textit{BeppoSAX} and \textit{INTEGRAL} observed the 62 X-ray burst sources. 

The energy spectra of thermonuclear bursts consist of emissions from the neutron star surface or boundary layer \citep{1989hanawa,2001popham,2003gilfanov}. Thermonuclear bursts can occur in two different spectral states, namely low/hard (island state) and high/soft (banana state) spectral states in the color-color diagram. The characteristic of the thermonuclear bursts is unique in both states. \citep{2008galloway}. The different spectral states are thought to originate from a different accretion process. The thermal instability triggers the viscous instability, as the temperature increases, the mass accretion rate also increases \citep{2007donec}. During the rising part of the burst, observed energy spectra are generally hard, whereas the spectrum becomes soft during the decay phase of the thermonuclear burst \citep{2011zhangg,2018dege}. In the high/soft state, disk approaches to the neutron star surface and terminate the emission from the boundary layer \citep{1984mitsud,2003gilf}. The disk is assumed to be geometrically thin and optically thick, while close to the central object. However, in low/hard state, the innermost accretion occurs through an optically thin along with the truncated disk and the emission from the source may originate from the hot corona \citep{1999uber,2007donec,2016degenaar,2017keekgal,2018bhatt}.

The energy spectrum of the thermonuclear burst sources is well-fit by a sum of blackbody components with interstellar absorption and a thermal Comptonization model. The blackbody component originates due to the emission from the innermost part of the accretion disk and the NS surface \citep{1977swank,2014keek,2016degenaar,2017galloway}. The measured blackbody temperature of X-ray bursts $kT_{bb}$ $\approx$ 1-3 keV that evolves with time \citep{2007donec,2017keekgal}. Moreover, several recent studies described the absorption and emission features observed during the X-ray bursts that might indicate to a mass point outflow or change in the accretion disk \citep{2004balla, 2005ballan, 2014bkeek, 2017galloway}. The luminosity and exposure of a burst depend on the mass accretion rate and chemical composition of the accreted material. At the peak, the burst emission may exceed the Eddington limit $L_{Edd}$ $\approx$ 2 $\times$ $10^{38}$ erg $s^{-1}$ \citep{1993lewin}.  During the expansion phase of the burst, the luminosity may reach or exceed the Eddington limit locally at the surface resulting in photospheric expansion. The photosphere regains its original size after the thermonuclear burning (rp-process) of H/He mixture ceases, and finally, the star cools down to roughly constant radius \citep{1993lewin,2004balla,2010boutl,2016degenaar,2017keekgal}.

4U 1323-62 is a faint LMXB source, which was first detected by UHURU and Ariel V \citep{1978forman,1981warwick}. The source exhibits regular thermonuclear type-1 X-ray bursts, and periodic intensity dips were first discovered by the ``EXOSAT" observations from February 11-13, 1985 \citep{1985vander,1989parmar,2016gambino}. MINBAR is used to locate the \textit{RXTE/PCA} observations of 4U 1323-62 that shows 40 bursts with a recurrence time of 2.45-2.59 hr.  From the characteristics of the bursts, it was constrained the distance of the source is 10 - 20 kpc \citep{1989parmar}. \cite{1987frank} suggested that the existence of periodic dips indicates the presence of high inclination angle ${\angle}$i = $60^{\circ}$ - $80^{\circ}$, where ${\angle}$i is the angle between the line of sight and the rotation axis of the accretion disk.  The source is located at $0.5^{\circ}$ above the galactic plane and shows a prominent dust halo \cite{2001barnard}. The well defined spectrum of this source was described by the \textit{BeppoSAX} and \textit{RXTE/PCA} observations \citep{1999balch,2009church}. In \textit{RXTE/PCA} observation, the author reports the presence of a Fe emission feature on a 6.43 keV with width 1.1 keV. In April 1997, a QPO was discovered at $\sim$ 1 Hz \citep{2010balman}. The QPO was detected in this source during the bursts, the dips, and as well as during the persistent emission \citep{1999jonker, 1999homann}.  
In this work, we present the detailed temporal and spectral study of the six thermonuclear bursts and two dips observed during     \textit{AstroSat/LAXPC} observation \ref{instruments1}. We have studied energy dependent burst profiles and time resolved spectroscopy of the source.

  \begin{table*}
   \begin{center}
   \begin{tabular} {c c c c c c c c}
 \hline 
Instrument & Obs. Time & Exposure (ks) & Bursts & Dips& Orbit Period & Reference \\
 \hline 
\textit{EXOSAT/ME} & February 11-13, 1985  & 108 & 6  & 10 & 2.932 hr  & \cite{1989parmar}  \\
\textit{RXTE/PCA} & April 25-28, 1997  & 200  & 7$^{**}$ & 7 & 2.45-2.59 hr   & \cite{2001barnard}  \\
\textit{BeppoSAX/MECS} & August 22 \& 24, 1997  & 120 & 10 & 12 & 2.40-2.57 hr     &  \cite{2009church}  \\
\textit{XMM-Newton} & January 29, 2003 & 50  &  7   & 5 &  2.97 hr &  \cite{2005chu} \\
\textit{RXTE/PCA} & September 25, 2003  & 200  & 7 & 4 & 2.45-2.59 hr   &  \cite{2016gambino}  \\
\textit{SUZAKU} & January 9-10, 2007  &   122.5   & 5$^*$ & 11 &  2.926 hr &  \cite{2009church}  \\
\textit{CHANDRA}  & December 19, 2011  & 60  &  9 & 6 & 2.94 hr & \cite{2016gambino} \\
\textit{CHANDRA} &  December 23, 2011 & 80   & 13$^*$ & 8 & 2.94 hr   & \cite{2016gambino}   \\
\textit{LAXPC} & February 16-17, 2017 & 49.5  & 6 & 2  & $\sim$2.66 hr & present \\
  
    \hline
    \end{tabular}
    \end{center}
   \caption{The observations logs of 4U 1323-62 data by different instruments detect the orbit period. \textbf{Symbol}$^*$ represent the detection of one doublet and $^{**}$ represent the two doublets observed in the light curve. }
    \label{instruments1}
    \end{table*}

\begin{figure}
\centering \includegraphics[width=0.35\textwidth,angle=-90]{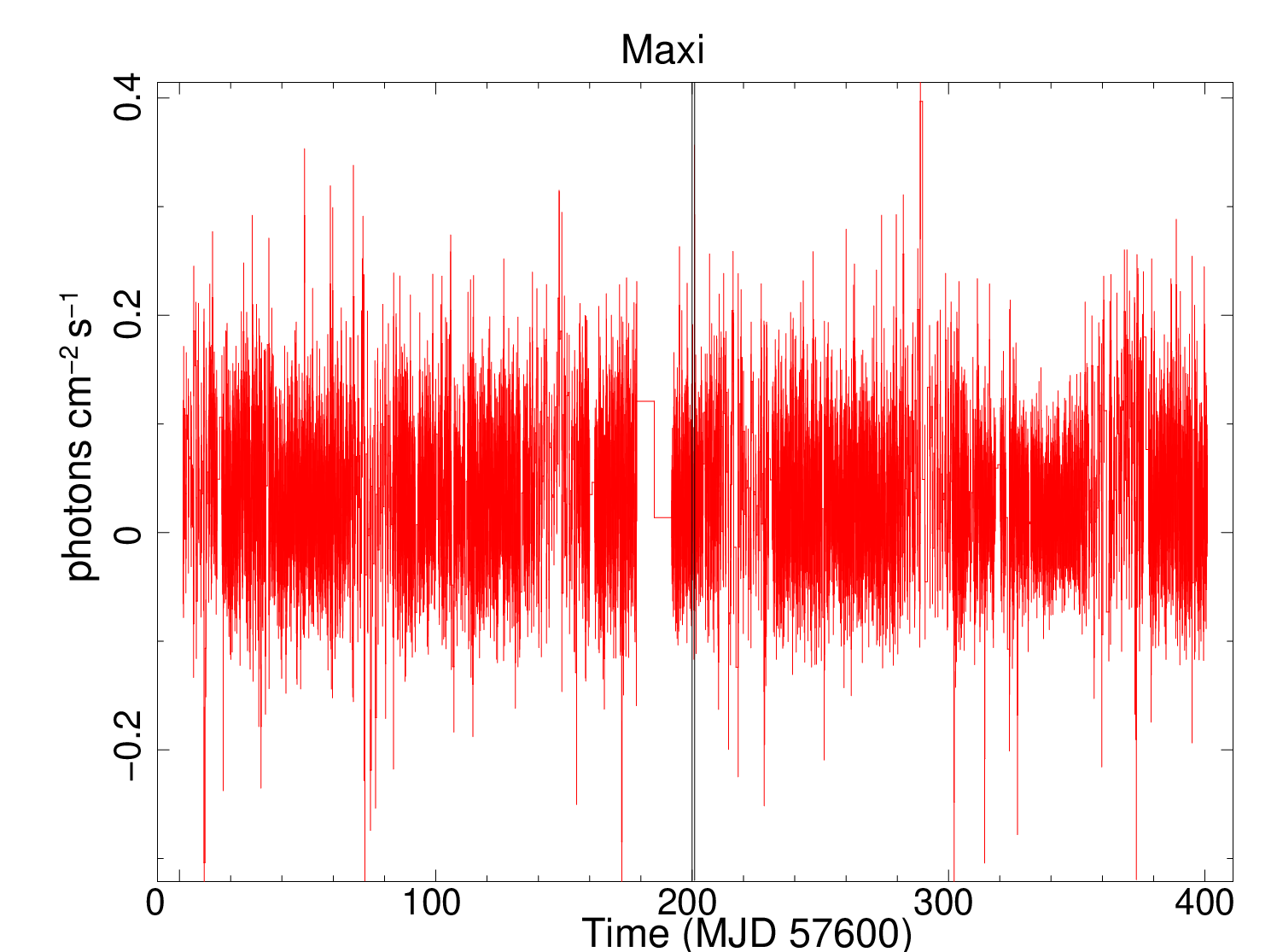}
\centering \includegraphics[width=0.35\textwidth,angle=-90]{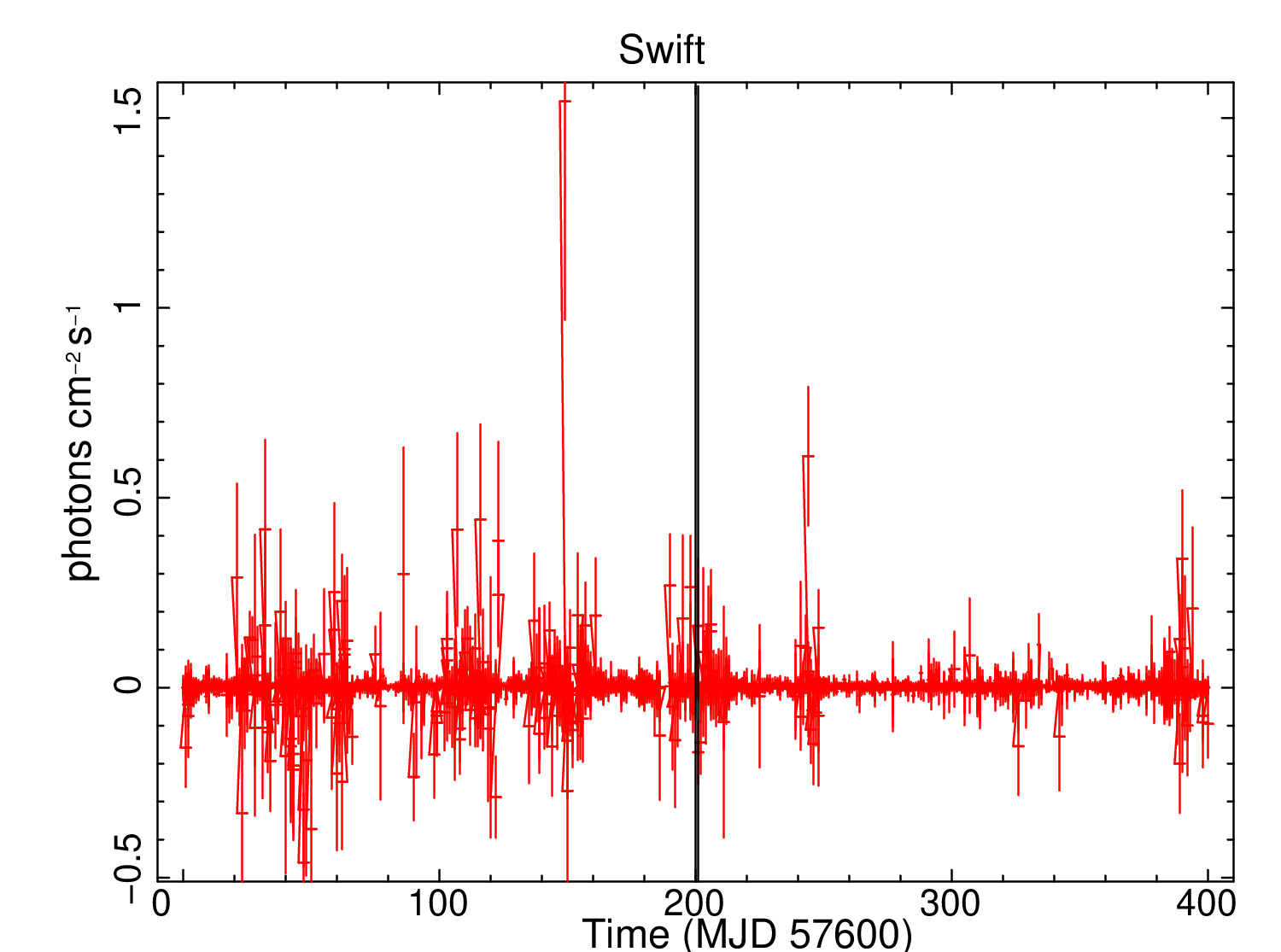}
\caption{These figures indicate the variability in the light curve of the 4U 1323-62 neutron star using the MAXI light curve in the 2-20 keV range in the left panel and Swift-BAT light curve in 15-50 keV range in the right panel. The bold line in both figures highlights the LAXPC observation time (i.e., MJD: 57800-57801).}
\label{maxi-swift_fig1}
\end{figure}

\begin{figure} 
\centering \includegraphics[width=0.60\textwidth,angle=-90]{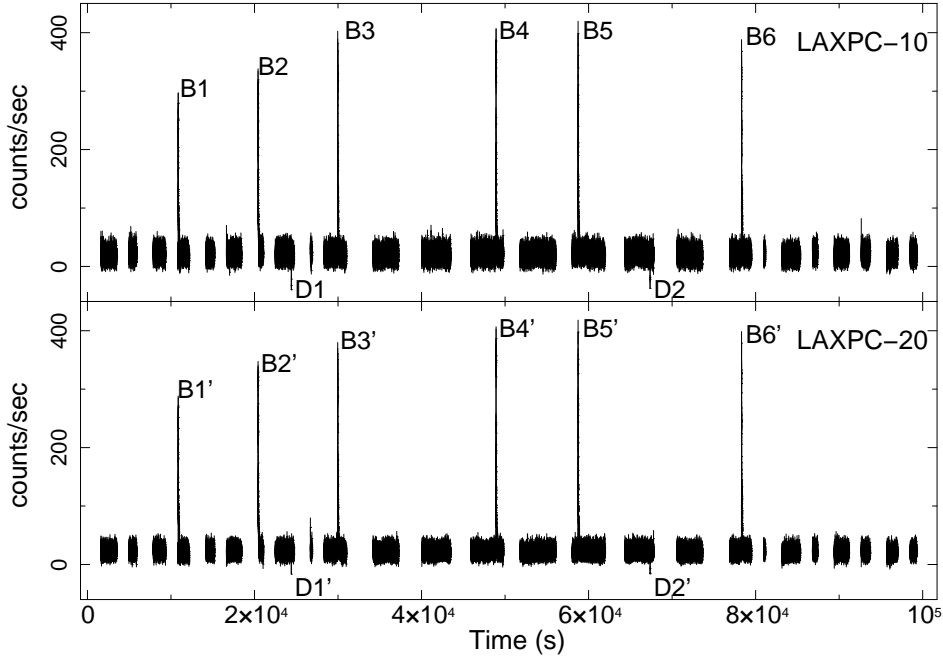}
\caption{Background subtracted light curve in 3-20 keV energy range using LAXPC-10 and LAXPC-20 data of 4U 1323-62. The light curve is binned with a bin size of 1 sec. Multiple bursts and dips have been shown in this figure. In the upper window of the plot, B and D stand for burst and dips, respectively, with consecutive numbers for LAXPC-10. B' and D' is the signature for burst and dips observed by LAXPC 20.}
\label{lc_fig2}
\end{figure}

\section{Observations and Data reduction}

\textit{AstroSat} is the first Indian multiwavelength space astronomy observatory launched on September 28, 2015. There are five instruments (Scanning Sky Monitor (SSM), Soft X-ray Telescope (SXT), Large Area X-ray Proportional Counter (LAXPC), Cadmium Zinc Telluride Imager (CZTI), Ultraviolet Imaging Telescope (UVIT)) onboard \textit{AstroSat} which cover a wide energy range \citep{2006agra,2013paulbb,2016yadavjs,2016Royj}. Large Area X-ray Proportional Counter (LAXPC) is one of the major instrument onboard \textit{AstroSat} Satellite. LAXPC instrument is providing high time resolution X-ray data in a 3-80 keV energy range. \textit{AstroSat} consist a set of three co-aligned identical LAXPC detector to provide total effective area of $\sim$ 6000 $cm^2$ at 15 keV \citep{2017antiahm}. The three LAXPC units is named as LAXPC-10, LAXPC-20, LAXPC-30.   The energy resolution of LAXPC units are 20\% and 10\% respectively at 6 and 60 keV, respectively \citep{2016yadav}. LAXPCs are a multi-anode and multi-layer configuration having a field of view of about $ 1^{\circ}$ $\times$ $1^{\circ}$. Each LAXPC unit has five layers consists of 12 detector cells. Each LAXPCs independently observes the source \citep{2016yadavjs,2017Agraw,2017antiahm}.

  \begin{table*}
   \begin{center}
   \begin{tabular} {c c c c c c c c}
 \hline 
& & & LAXPC-10 & &  &\\
\hline
 Burst  & Start-End Time & Rising Time (t') & Decay Time (t") & Burst Exposure & Wait Time  & Peak Intensity \\
  &  & (sec.) & (sec.) & (sec.) & (hour) & (count/sec) \\
B1 & 10856-10931 &13  & 62   &  75  & - & 295 \\
B2 &20423-20513 &6  & 84   & 90   & 2.636 & 318 \\
B3  &29962-30056&8  & 86   & 94   & 2.625 & 382 \\
B4  &48914-49015 & 11 &  90  &  101  & ----- & 388  \\
B5 &  58737-58830 &10  & 83   &  93  & 2.70 & 399 \\
B6 & 78312-78410 & 11 &  87  &  98  & ----- & 368 \\
 \hline
 Dip  & Start-End Time & Rising Time (t') & Decay Time (t") & Burst Exposure & Wait Time   \\
 &  & (sec.) & (sec.) & (sec.) & (hour)  \\
D1  & 24416-24421 & 3 & 2 & 5 & --- \\ 
D2 & 67366-67370 & 2 & 2 & 4  & 11.928  \\
 \hline 
& & & LAXPC-20 & &  &\\
\hline
 Burst  & Start-End Time & Rising Time (t') & Decay Time (t") & Burst Exposure & Wait Time  & Peak Intensity \\
  &  & (sec.) & (sec.) & (sec.) & (hour) & (count/sec) \\
B1' & 10857-10932.5 & 13  & 62.5   &  75.5  & - & 289 \\
B2' &20423-20512 & 7  & 82  & 89   & 2.636 & 328 \\
B3'  &29962-30057& 9  & 86   & 95   & 2.625 & 367 \\
B4'  &48914-49013 & 11 & 88  &  99  & ----- & 390  \\
B5' &  58737-58829 & 9  & 83   &  92  & 2.70 & 398 \\
B6' & 78312-78411 & 12 &  87  &  99  & ----- & 378 \\
 \hline
 Dip  & Start-End Time & Rising Time (t') & Decay Time (t") & Burst Exposure & Wait Time   \\
 &  & (sec.) & (sec.) & (sec.) & (hour)  \\
D1'  & 24416.5-24419 & 1 & 2 & 3 & --- \\ 
D2' & 67366-67370 & 2 & 2 & 4  & 11.928  \\

    \hline
    \end{tabular}
    \end{center}
   \caption{The X-ray bursts and dips, rising and decay time interval, along with wait time observed using LAXPC 10 and LAXPC 20, are tabulated.}.
    \label{time}
    \end{table*}

LAXPCs collect the data in two different modes: Event Analysis mode (modeEA) and Broad Band Counting (modeBB). Event mode data give information about the energy of each photon and its arrival time at 10 micro-sec accuracies. We have used event analysis mode modeEA data for generating the light curves, power density spectrum (PDS) and energy spectrum of the source. Due to the wide energy range and large photon collecting area, \textit{AstroSat}, efficiently observes the thermonuclear X-ray burst sources. The source 4U 1323-62 was observed by \textit{AstroSat/LAXPC} from February 16, 2017 05:25:43 till February 17, 2017 08:35:35 for total effective exposure time 49.5 ks. The LAXPC data were downloaded from the ASSC astrobrowser\footnote{https://astrobrowse.issdc.gov.in/astro$_{-}$archive/archive}.

  \begin{figure*}
\centering \includegraphics[width=0.32\textwidth,angle=-90]{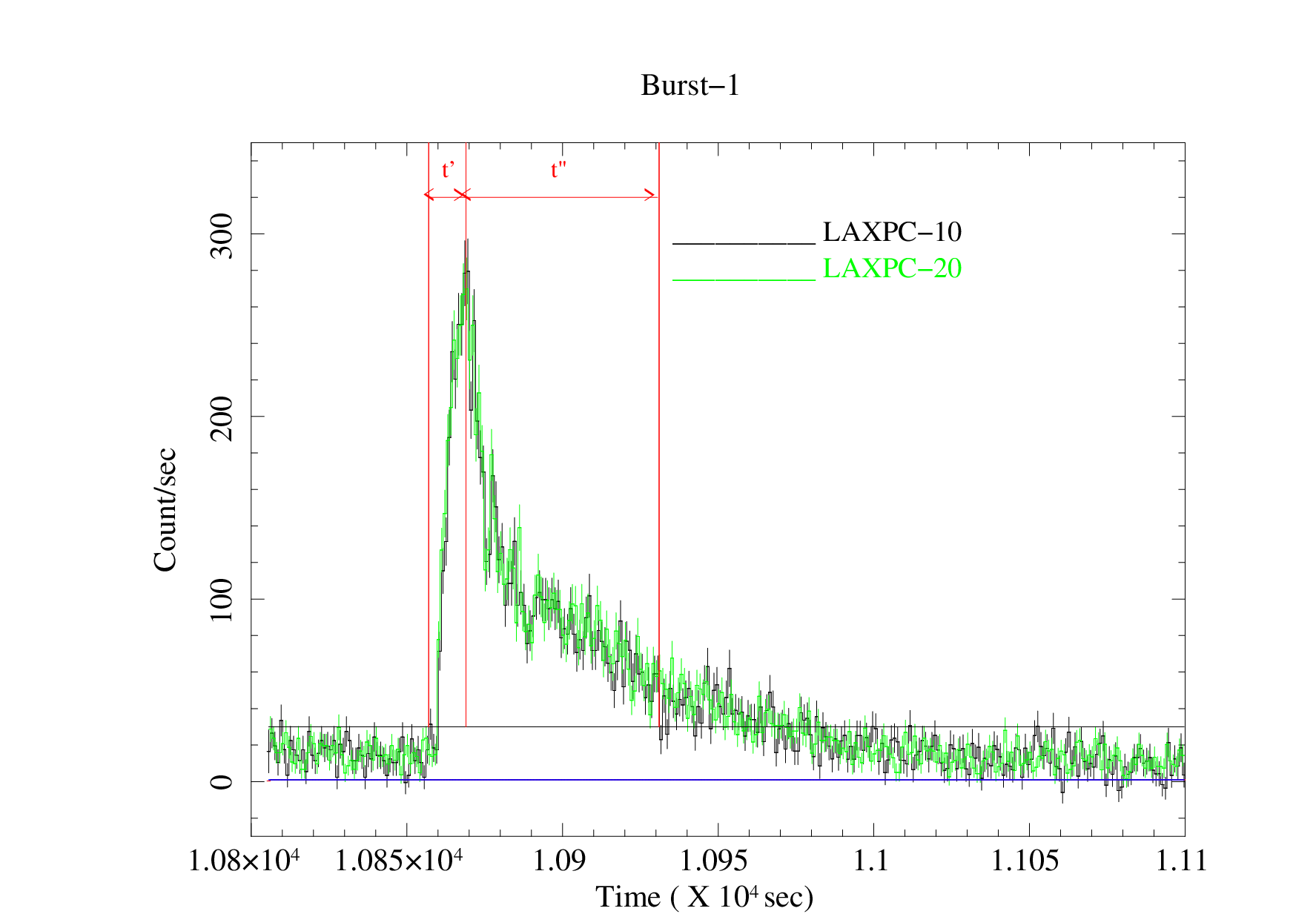}
\centering \includegraphics[width=0.32\textwidth,angle=-90]{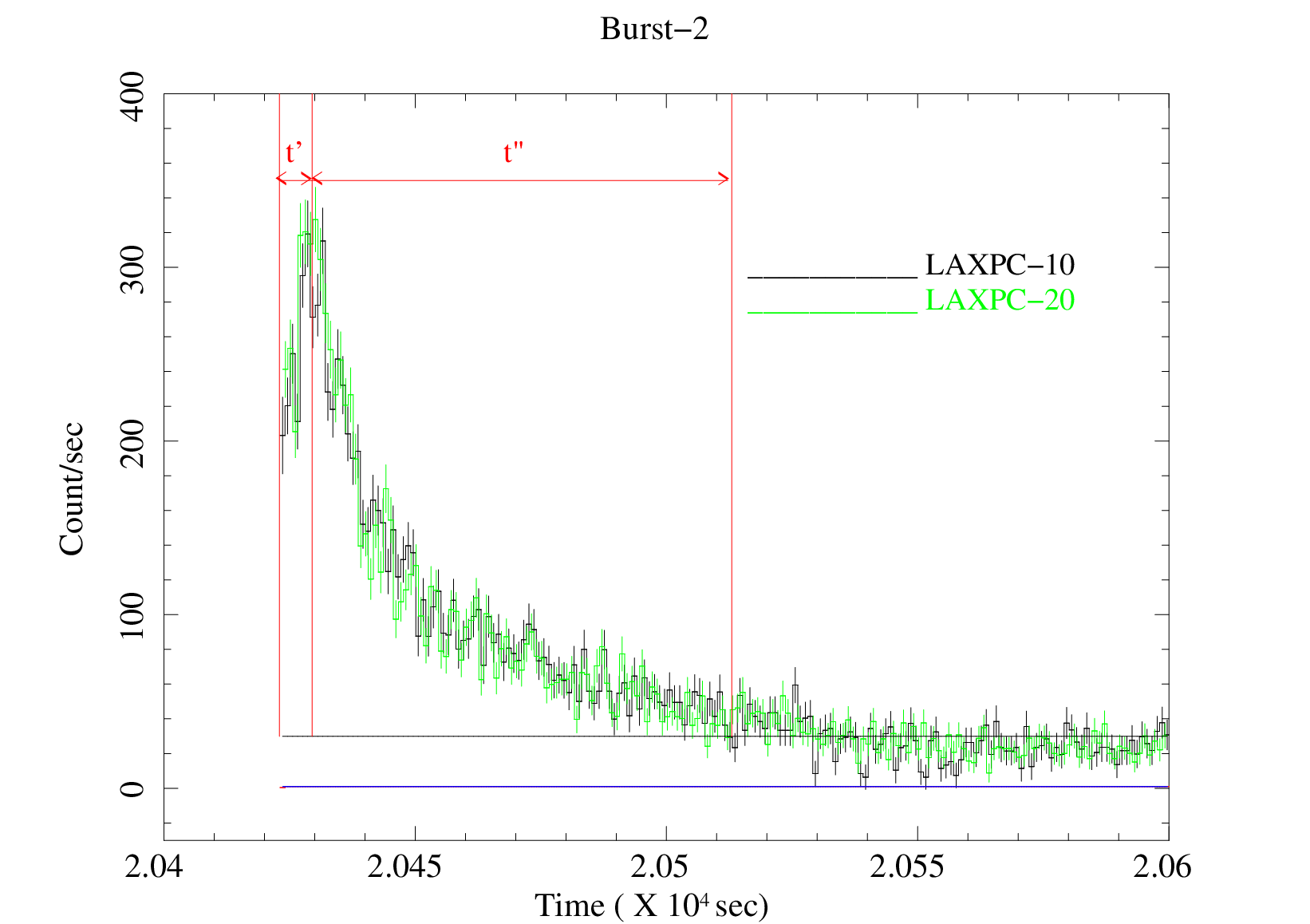}
\centering \includegraphics[width=0.32\textwidth,angle=-90]{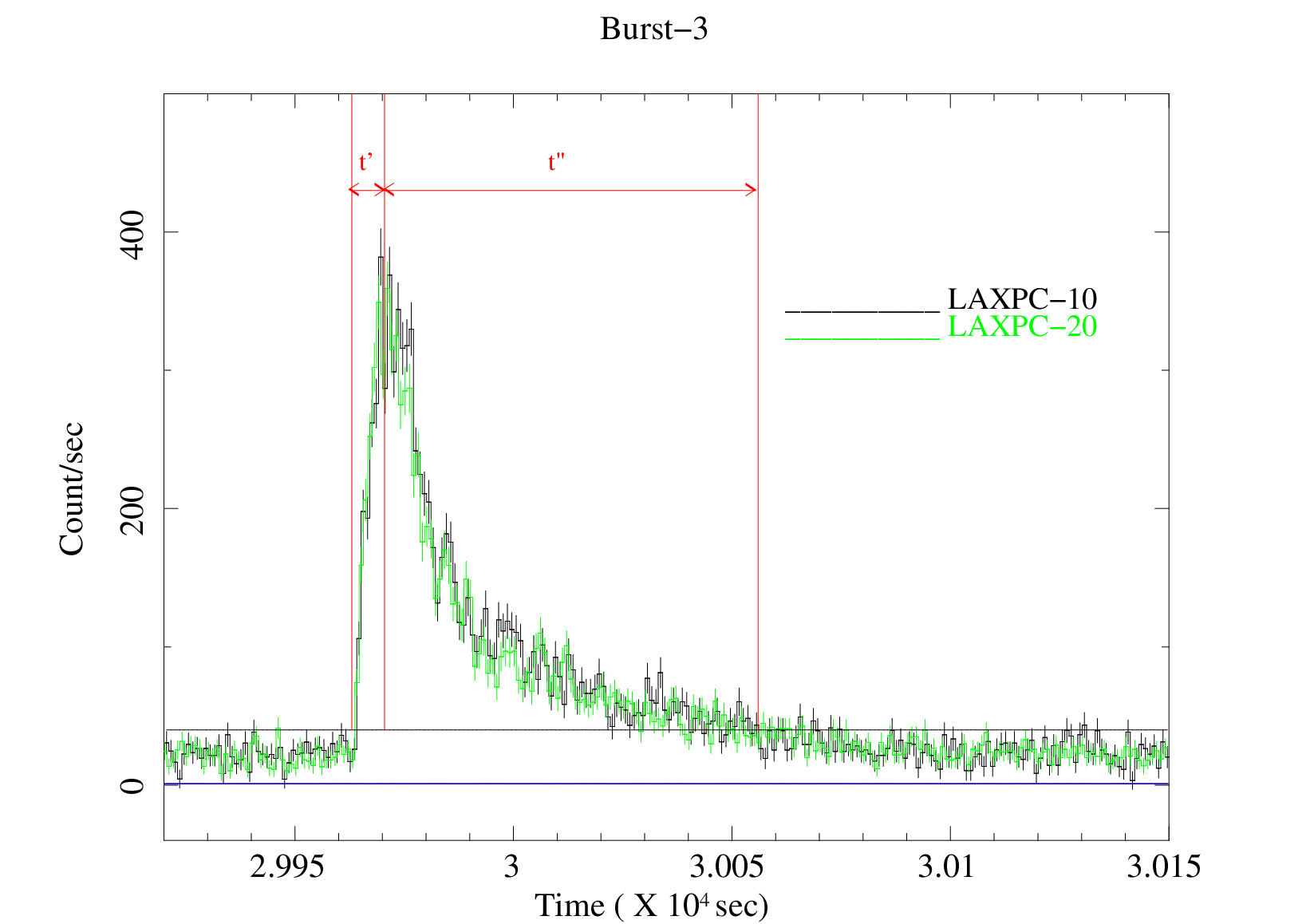}
\centering \includegraphics[width=0.32\textwidth,angle=-90]{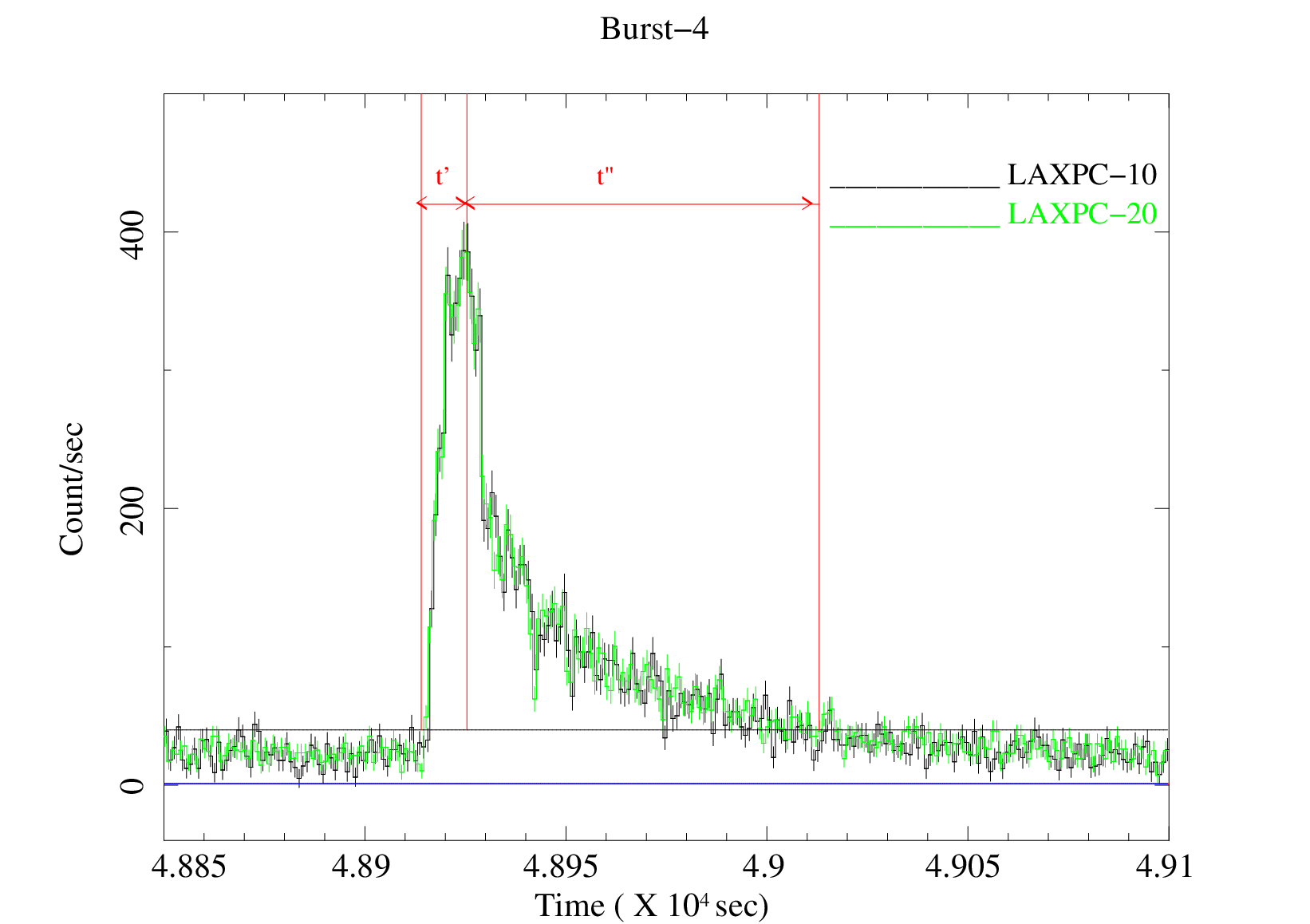}
\centering \includegraphics[width=0.32\textwidth,angle=-90]{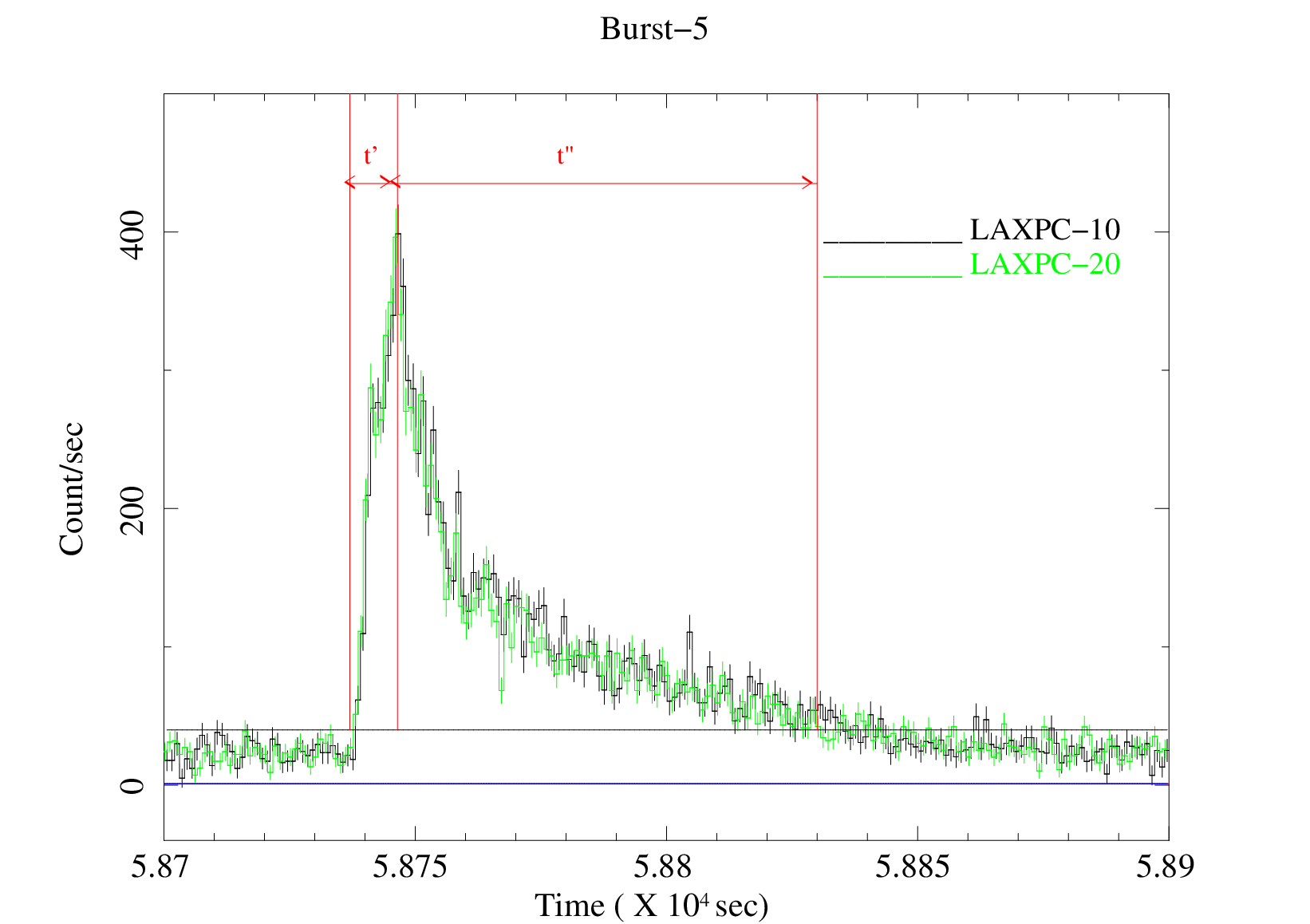}
\centering \includegraphics[width=0.32\textwidth,angle=-90]{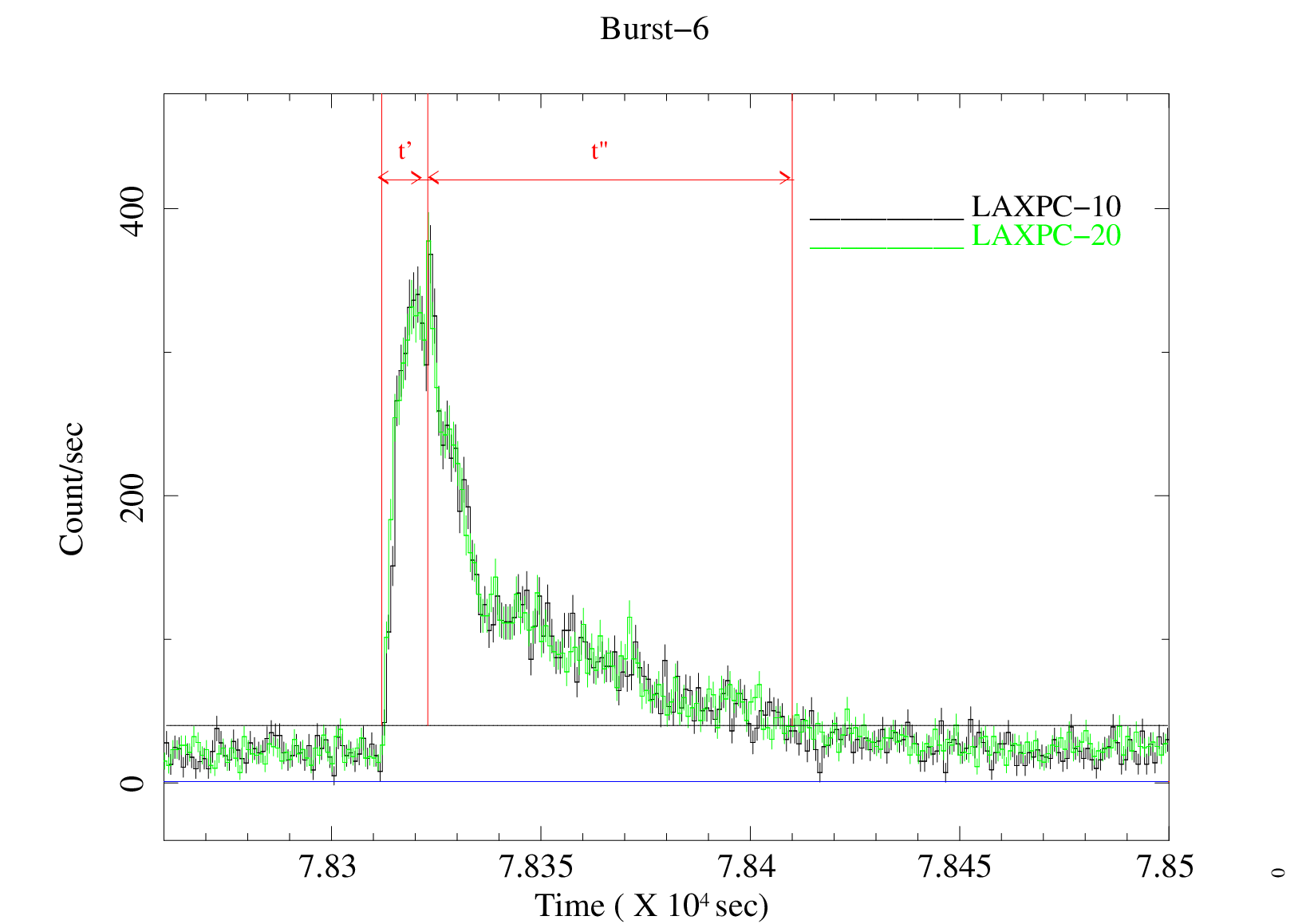}

\caption{The figures show the  bursts and dips in 3-20 keV energy range using LAXPC-10 (black) and LAXPC-20 (green) observations in 1 second binned light curve.}
\label{pcu_12_fig3}
\end{figure*}

 \begin{figure*}
\centering \includegraphics[width=0.32\textwidth,angle=-90]{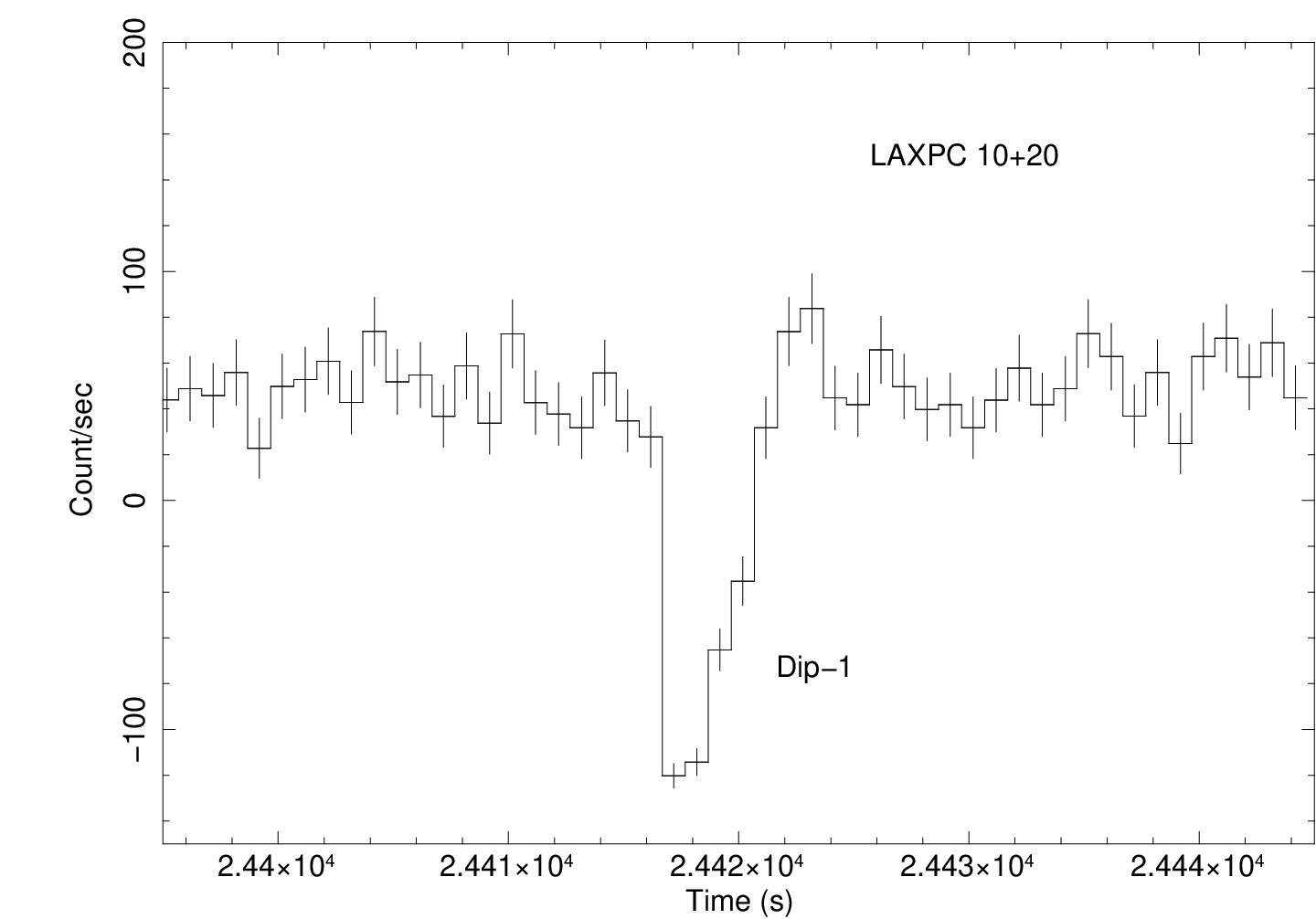}
\centering \includegraphics[width=0.32\textwidth,angle=-90]{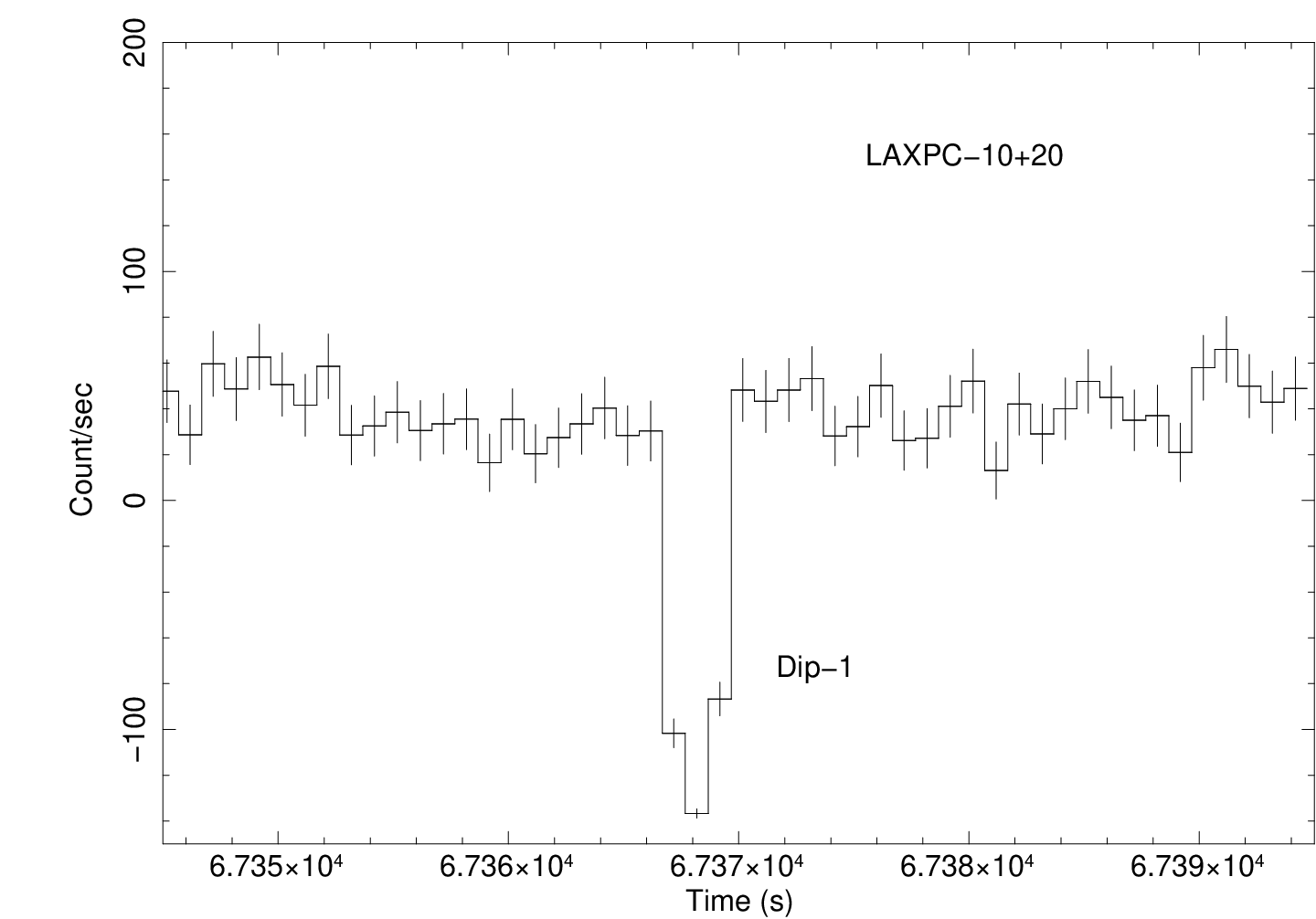}

\caption{The figures show the dips in 3-20 keV energy range using merge LAXPC-10 and LAXPC -20 observations in 1 second binned light curve.}
\label{merge-dip_fig4}
\end{figure*}

\section{Data Analysis and Results}

The data are analyzed using the individual routine LaxpcSoft\footnote{http://AstroSat-ssc.iucaa.in/?q=laxpcData} to obtain background subtracted light curves and energy spectra \citep{2017antiahm}. The commands are used to extract the light curve, and the spectral file is $`laxpc_{-}make_{-}lightcurve'$ and  $`laxpc_{-}make_{-}spectra'$. The background files for light curve and spectra are extracted by using the commands $`laxpc_{-}make_{-}backlightcurve'$ and  $`laxpc_{-}make_{-}backspectra'$. The data analysis was done by the HEASOFT 6.19\footnote{http://heasarc.nasa.gov/ftools}, which consists mainly of FTOOLS\footnote{https://heasarc.gsfc.nasa.gov/ftools}, XORONOS and XSPEC \citep{1996arnaudka,2008noblems}. We observed that the thermonuclear bursts and dips in the lightcurve are present in both LAXPC units (LAXPC 10 and 20), as shown in figure \ref{lc_fig2}. For generating the light curves and spectral analysis, we used data only from the top layer of LAXPC-10 and LAXPC-20. The top layer of the LAXPC detector absorbs 90$\%$ of the X-ray photons below 20 keV \citep{2016Royj}. We have not used the data of LAXPC-30 for this work due to gain instability caused by gas leakage, as suggested by \cite{2017antiahm}.

The results of previously reported thermonuclear bursts and dips observed from the source from various missions are presented in Table \ref{instruments1}. This table indicates the reports of orbital period estimates based on the recurrence time of the thermonuclear bursts. The light curves for each burst and dips for both of LAXPC detector, as shown in figure \ref{pcu_12_fig3} and \ref{merge-dip_fig4} respectively, where the horizontal and vertical lines indicate the rise (t') and decay (t'') time of the bursts.

\subsection{Light curves}

We used the orbit binned light curve of MAXI\footnote{http://maxi.riken.jp/star$_{-}$data/J1326-621/J1326-621.html} transient monitor in the soft energy band (3-20 keV) and the Swift/BAT transient Monitor \citep{2013Krimm} lightcurve in the hard energy band (15-50 keV) of 4U 1323-62 to determine the spectral state of this source. In both instruments, the light curve of 4U 1323-62 shows the long-term variability in intensity from  July 31, 2016, to  September 04, 2017. The bold line in both figures \ref{maxi-swift_fig1} highlights the LAXPC observation time of 4U 1323-62. In the figure \ref{lc_fig2}, LAXPC-10 and LAXPC-20 observations of 4U 1323-62 show the light curve in 3-20 keV energy range binned with a bin size of 1 second. In light curve data, gaps and thermonuclear X-ray bursts are detected.

  \begin{table*}
   \begin{center}
   \begin{tabular} {c c c c c c c c}

\hline

Energy & Peak-count & Burst-rising & Burst-decay &    \\
keV & count/s & sec. & sec.     \\
3-6.2 & 140  & 9.5  &  84.5 \\
6.2-9 & 119 & 10 & 85.5  \\
9-12 & 72 & 10 & 91 \\
12-16 & 47 & 10.5 & 96 \\
16-20 & 24 & 18 & 22\\

    \hline
    \end{tabular}
    \end{center}
   \caption{This table shows the maximum count rate, rising and decaying time of fifth thermonuclear X-ray burst in different energy bands.}
    \label{ers_count}
    \end{table*}

During the 100 ks observation of AstroSat/LAXPC observation of the source, six thermonuclear bursts are observed, as shown in Figure 2. We have used nomenclature of B1, B2, B3, B4, B5, and B6 to indicate the thermonuclear bursts in the LAXPC 10 lightcurve and B1', B2', B3', B4', B5' and B6' in the LAXPC 20 lightcurve respectively. We observed that there is a gap between the successive bursts, B3 and B4, and the bursts B5 to B6 are twice the wait time observed during the consecutive bursts B1 and B2. These missed bursts are attributed to data gaps. The data gaps, during the occurrence of bursts, are due to the position of the satellite on the South Atlantic Anomaly (SAA) region.  

Each burst has an exposure time (rising and decay time) between 75-100 sec and a high peak count rate of $\approx$ 380 count/sec, as shown in the light curve (Figure \ref{pcu_12_fig3}). In Table \ref{time}, the burst rising and decay time interval, exposure of each burst, wait time, and the peak count rate have been tabulated. In figure \ref{merge-dip_fig4} prominent dips have been shown in 3-20 keV energy range using observations of both LAXPCs unit (LAXPC 10 and LAXPC 20).

\subsection{Energy-Resolved Burst Profiles}

\begin{figure}
\centering \includegraphics[width=0.50\textwidth,angle=-90]{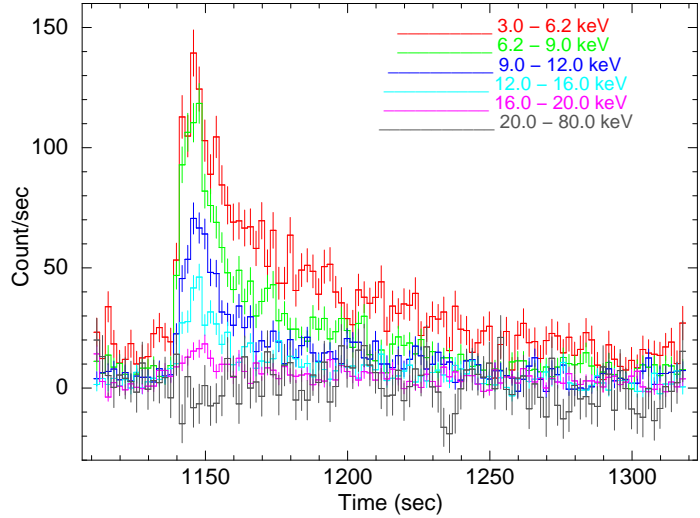}                 
\caption{This plot shows a light curve of the fifth thermonuclear X-ray burst in different energy bands. The softest energy range (3-6.2 keV) has the highest count rate.}
\label{bursts-fifth_fig5} 
\end{figure}

To probe the thermonuclear X-ray burst in different energy bands, we generated 1sec binned light curves from LAXPC 10 and 20 during an X-ray burst in six different narrow energy bands: 3-6.2 keV, 6.2-9 keV, 9-12 keV, 12-16 keV, 16-20 keV, and 20-80 keV. The light curves were generated from only the top layer of LAXPC 10 and LAXPC 20. Figure \ref{bursts-fifth_fig5} shows the energy-resolved light curve of the fifth burst (B5) seen in the observation of 4U 1323-62. The energy dependent thermonuclear burst profile clearly showed that the decay time decreases with the increase in energy.
  
As per table \ref{ers_count}, the maximum intensity is accumulated at a low energy range (3-6.2 keV). For the soft energy region, the rising time is relatively shortest. However, the energy range from 16-20 keV the peak intensity is lowest, and the rising time took maximum time to reach a peak intensity level, and in 20-80 keV energy range, the thermonuclear burst is absent. It may be inferred from the figure \ref{bursts-fifth_fig5} and table \ref{ers_count}, that 4U 1323-62 was observed in a soft spectral state during the AstroSat/LAXPC observations.

We measured the energy dependence of the burst duration in all the six X-ray bursts. We found that there was a gradual decay in the intensity with an increase in energy. The rising and decay time of all the bursts have been tabulated in Table \ref{time}.

\subsection{Power Density Spectra}

\begin{figure}
\centering \includegraphics[width=0.36\textwidth,angle=-90]{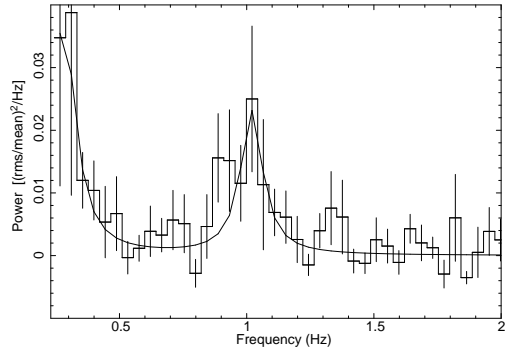}
\caption{Power density spectra generated from 0.01 sec binned light curve by merging LAXPC 10 and 20, show the low frequency $\sim$ 1 Hz QPO in the 3-20 keV energy range.}
\label{pds_fig6} 
\end{figure}

In the persistent emission, PDS was computed in the 3-20 keV energy band. The light curve extracted from  LAXPC-10 and 20 was used to generate PDS using the `FTOOLS' task ``Powspec''. The Poissonian noise level is subtracted from each power spectra. PDS  was Leahy normalized ($(rms/mean)^2/Hz$), such that the integration of the power spectra gave the fractional root mean square (r.m.s) variation.

A clear QPO was detected in the power spectra during the bursts. The QPO was fitted using a Lorentzian model. We have observed the low frequency QPO $\sim$ 1 Hz, which was reported by \cite{2010balman} using RXTE/PCA data. To detect the LFQPO, we used a light curve binned with a bin size of 0.01 sec. The light curve was divided into each segment with 2048 bins.
Figure \ref{pds_fig6} shows the QPO detected during the thermonuclear burst. The observed $\sim$ 3 sigma significant QPO frequency is 0.97 Hz. The r.m.s of the QPO = 6.7\%, and the quality factor is $\sim$ 18. \cite{1999jonker} previously discovered a low frequency 0.77-0.87 Hz QPO from the source 4U 1323-62 using RXTE/PCA data. The amplitude of the QPO detected from the source was found to be constant during the persistent emission, X-ray dips during the type-1 X-ray thermonuclear bursts \citep{1999jonker}.

\begin{figure*}
\centering \includegraphics[width=0.50\textwidth,angle=-90]{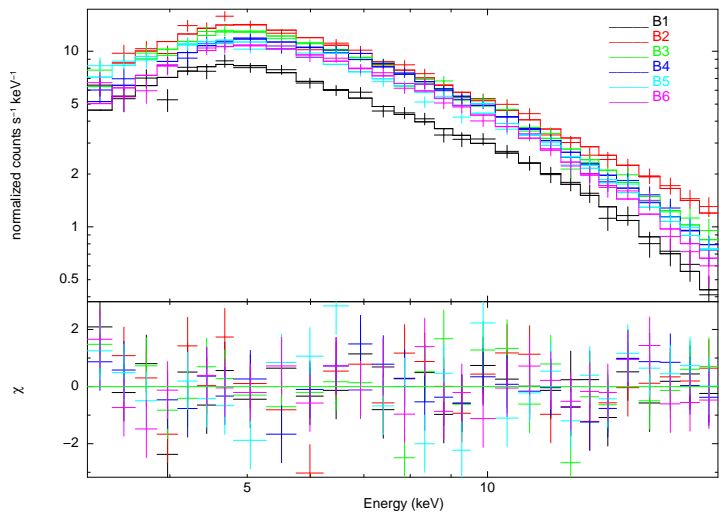}
\caption{The energy spectra of the six thermonuclear X-ray bursts observed by LAXPC 10 in the 3-20 keV energy range is shown in this figure.}
\label{spectra-all_fig7} 
\end{figure*}

 \begin{figure*}
\centering \includegraphics[width=0.35\textwidth,angle=-90]{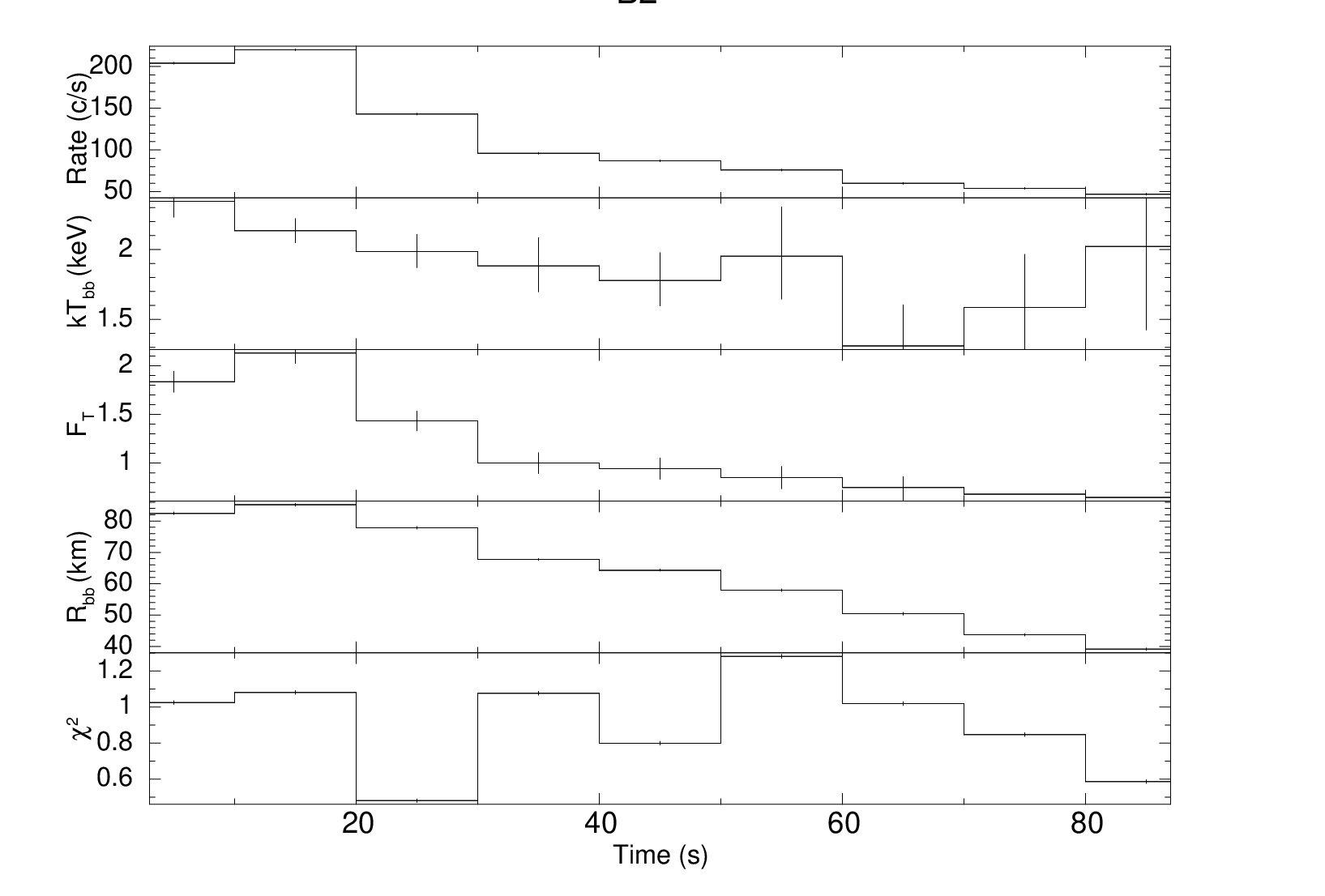}
\centering \includegraphics[width=0.35\textwidth,angle=-90]{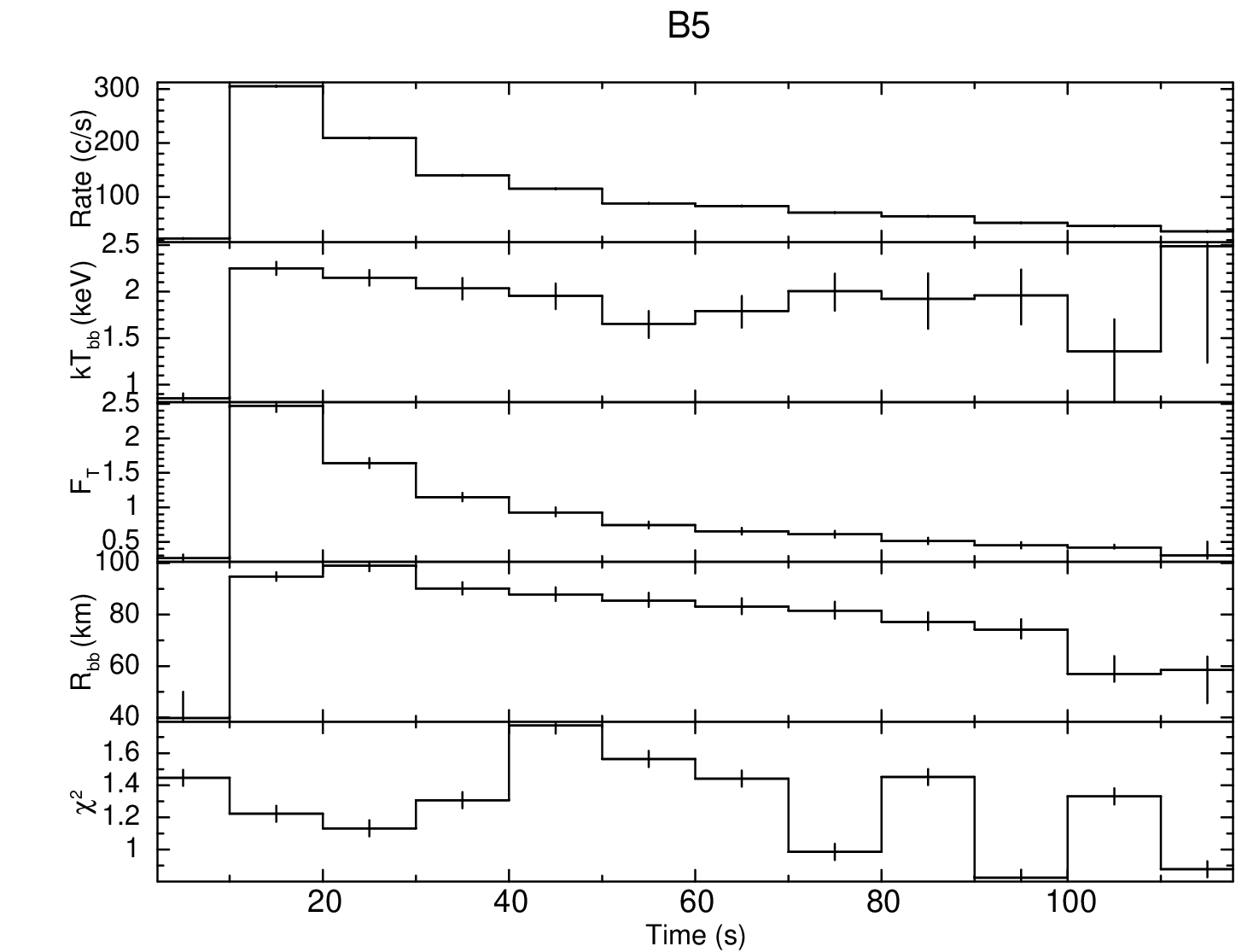}
\caption{The best-fit parameters obtained from performing time-resolved spectroscopy of two of the observed bursts (B2 and B5) in LAXPC-10 observation is shown in the figure. The top panel shows the average count rate (without error) of each segment.}
\label{d1-tbabs}
\label{multiple_fig8}
\end{figure*}

\subsection{Energy Spectra : Time resolved spectroscopy}

Spectral fitting for individual X-ray bursts was carried out. To investigate the spectral evolution during X-ray bursts, we performed time-resolved spectroscopy of each type-1 thermonuclear bursts. Therefore the bursts data were divided into segments of 10 seconds intervals. Each burst has 8-12 segments. These 10s data segments during each burst are used to reveal the evolution of the blackbody component in energy spectra. For each burst excepts the second burst (B2), we extracted the spectra for 10  seconds of data preceding the bursts. In the burst B2, we have extracted the spectra from data after the burst decayed completely to the persistent levels. There was no difference in the distributions of best-fitting persistent models between the pre- and post-bursts.  The spectra extracted from persistent emission was used to subtract for all bursts segments as the background.

  \begin{table*}
   \begin{center}
   \begin{tabular} {c c c c c c c c}
 \hline 
& & LAXPC-10 & & \\
\hline
&  &  Burst-1  &  & \\
\hline
 Segments  & Rate (count/s) & $kT_{in}$ (keV)  & Flux $\times$ $10^{-9}$ & Radius (km)  & Chi-square \\
\hline
$5^{+5}_{-5}$ & $15^{+1}_{-1}$ & $3.05^{+0.27}_{-0.32}$ & $0.11^{+0.035}_{-0.050}$ & $74.31^{+5.94}_{-5.94}$ & $1.074^{+0.05}_{-0.05}$ \\
$15^{+5}_{-5}$ & $183^{+1}_{-1}$ & $2.19^{0.092}_{-0.096}$ &  $1.47^{+0.067}_{-0.068}$ & $95.62^{+2.42}_{-2.15}$ & $0.91^{+0.05}_{-0.05}$ \\
$25^{+5}_{-5}$ & $179^{+1}_{-1}$ & $2.07^{+0.090}_{-0.094}$ & $1.48^{+0.056}_{-0.076}$ & $96.00^{+2.21}_{-2.25}$& $1.34^{+0.05}_{-0.05}$  \\
$35^{+5}_{-5}$ & $109^{+1}_{-1}$ & $1.89^{+0.11}_{-0.12}$ & $ 0.81^{+0.05}_{-0.05}$ & $93.82^{+3.03}_{-2.12}$ & $1.73^{+0.05}_{-0.05}$  \\
$45^{+5}_{-5}$ & $98^{+1}_{-1}$ & $1.79^{+0.11}_{-0.12}$ & $0.74^{+0.048}_{-0.048}$ & $93.56^{+3.16}_{-2.44}$  & $1.08^{+0.05}_{-0.05}$  \\
$55^{+5}_{-5}$ & $82^{+1}_{-1}$ & $1.76^{+0.14}_{-0.16}$ & $0.62^{+0.044}_{-0.045}$ & $92.32^{+3.48}_{-3.05}$ & $1.18^{+0.05}_{-0.05}$  \\
$65^{+5}_{-5}$ & $68^{+1}_{-1}$ & $1.75^{+0.16}_{-0.18}$ & $0.53^{+0.042}_{-0.042}$ & $90.87^{+3.87}_{-4.22}$ & $0.92^{+0.05}_{-0.05}$  \\
$75^{+5}_{-5}$ & $61^{+1}_{-1}$ & $1.7^{+0.19}_{-0.22}$ & $0.41^{+0.039}_{-0.039}$ & $88.37^{+4.46}_{-4.39}$ & $1.97^{+0.05}_{-0.05}$  \\
$85^{+5}_{-5}$ & $41^{+1}_{-1}$ & $1.44^{+0.20}_{-0.24}$ & $0.31^{+0.036}_{-0.037}$ & $85.19^{+5.46}_{-3.09}$ & $1.21^{+0.05}_{-0.05}$  \\ 
$95^{+5}_{-5}$ & $43^{+1}_{-1}$ & $1.24^{+0.25}_{-0.30}$ & $0.50^{+0.06}_{-0.036}$ & $67.74^{+5.02}_{-3.8}$ & $1.68^{+0.05}_{-0.05}$ \\ 
$105^{+5}_{-5}$ & $42^{+1}_{-1}$ & $1.55^{+0.26}_{-0.31}$ & $0.29^{+0.035}_{-0.035}$ & $83.57^{+5.47}_{-4.77}$ & $0.95^{+0.05}_{-0.05}$ \\

\hline
& &  Burst-2  &   &  \\
\hline
 Segments  & Rate (count/s) & $kT_{in}$ (keV)  & Flux $\times$ $10^{-9}$ & Radius (km)  & Chi-square \\
\hline
$5^{+5}_{-5}$ & $204^{+1}_{-1}$ & $2.34^{+0.11}_{-0.11}$ & $1.83^{+0.10}_{-0.10}$ & $82.44^{+0.43}_{-0.37}$ & $1.02^{+0.01}_{-0.01}$ \\
$15^{+5}_{-5}$ & $220^{+1}_{-1}$ & $2.13^{+0.087}_{-0.084}$ & $2.13^{+0.10}_{-0.10}$ & $85.20^{+0.37}_{-0.46}$ & $1.08^{+0.01}_{-0.01}$ \\
$25^{+5}_{-5}$ & $143^{+1}_{-1}$ & $1.98^{+0.12}_{-0.11}$ & $1.43^{+0.10}_{-0.10}$ & $77.81^{+0.43}_{-0.34}$ & $0.47^{+0.01}_{-0.01}$ \\
$35^{+5}_{-5}$ & $96^{+1}_{-1}$ & $1.88^{+0.20}_{-0.18}$ & $1.0^{+0.10}_{-0.10}$ & $67.73^{+0.31}_{-0.31}$ & $1.07^{+0.01}_{-0.01}$ \\
$45^{+5}_{-5}$ & $87^{+1}_{-1}$ & $1.77^{+0.19}_{-0.18}$ & $0.94^{+0.10}_{-0.10}$ & $64.30^{+0.33}_{-0.25}$ & $0.79^{+0.01}_{-0.01}$ \\
$55^{+5}_{-5}$ & $76^{+1}_{-1}$ & $1.95^{+0.35}_{-0.30}$ & $0.85^{+0.11}_{-0.11}$ & $57.95^{+0.37}_{-0.38}$ & $1.28^{+0.01}_{-0.01}$ \\
$65^{+5}_{-5}$ & $60^{+1}_{-1}$ & $1.31^{+0.29}_{-0.30}$ & $0.74^{+0.11}_{-0.15}$ & $50.45^{+0.38}_{-0.43}$ & $1.02^{+0.01}_{-0.01}$ \\
$75^{+5}_{-5}$ & $54^{+1}_{-1}$ & $1.58^{+0.37}_{-0.48}$ & $0.68^{+0.49}_{-0.19}$ & $43.75^{+0.31}_{-0.36}$ & $0.84^{+0.01}_{-0.01}$ \\
$85^{+5}_{-5}$ & $47^{+1}_{-1}$ & $2.02^{+0.56}_{-0.59}$ & $0.64^{+0.46}_{-0.17}$ & $39.18^{+0.31}_{-0.37}$ & $0.59^{+0.01}_{-0.01}$ \\
\hline
\hline
& & Burst-3 &    &  \\
\hline
 Segments  & Rate (count/s) & $kT_{in}$ (keV)  & Flux $\times$ $10^{-9}$ & Radius (km)  & Chi-square \\
\hline

$5^{+5}_{-5}$ & $26^{+1}_{-1}$ & $0.77^{+0.23}_{-0.30}$ & $ 0.23^{+0.052}_{-0.046}$ & $75.53^{+6.73}_{-5.15}$ & $0.935^{+0.05}_{-0.05}$ \\
$15^{+5}_{-5}$ &$317^{+1}_{-1}$ & $2.24^{+0.070}_{-0.072}$ & $ 2.30^{+0.083}_{-0.083}$ & $95.16^{+1.77}_{-1.69}$ & $1.33^{+0.05}_{-0.05}$ \\
$25^{+5}_{-5}$ & $193^{+1}_{-1}$ & $2.21^{+0.078}_{-0.081}$ & $0.20^{+0.012}_{-0.10}$ & $93.85^{+2.62}_{-2.12}$ & $1.12^{+0.05}_{-0.05}$ \\
$35^{+5}_{-5}$ & $116^{+1}_{-1}$ & $1.99^{+0.11}_{-0.12}$ & $0.10^{+0.11}_{-0.11}$ & $91.32^{+2.22}_{-2.7}$ & $1.16^{+0.05}_{-0.05}$ \\
$45^{+5}_{-5}$ & $102^{+1}_{-1}$ & $1.79^{+0.12}_{-0.13}$ & $0.80^{+0.050}_{-0.050}$ & $88.84^{+2.92}_{-2.66}$ & $1.098^{+0.05}_{-0.05}$ \\
$55^{+5}_{-5}$ & $80^{+1}_{-1}$ & $1.82^{+0.16}_{-0.18}$ & $0.65^{+0.047}_{-0.047}$ & $85.90^{+3.27}_{-3.75}$ & $1.15^{+0.05}_{-0.05}$ \\
$65^{+5}_{-5}$ & $65^{+1}_{-1}$ & $1.68^{+0.18}_{-0.19}$ & $0.49^{+0.043}_{-0.043}$ & $81.62^{+3.87}_{-3.05}$ & $1.77^{+0.05}_{-0.05}$ \\
$75^{+5}_{-5}$ & $62^{+1}_{-1}$ & $1.64^{+0.20}_{-0.22}$ & $0.45^{+0.044}_{-0.038}$ & $78.25^{+2.8}_{-3.33}$ & $1.11^{+0.05}_{-0.05}$ \\
$85^{+5}_{-5}$ & $50^{+1}_{-1}$ & $1.59^{+0.21}_{-0.25}$ & $0.40^{+0.040}_{-0.040}$ & $76.77^{+3.88}_{-4.24}$ & $1.14^{+0.05}_{-0.05}$ \\
$95^{+5}_{-5}$ & $44^{+1}_{-1}$ & $1.39^{+0.24}_{-0.30}$ & $0.36^{+0.043}_{-0.041}$ & $73.89^{+3.06}_{-3.77}$ & $1.02^{+0.05}_{-0.05}$ \\
$105^{+5}_{-5}$ & $33^{+1}_{-1}$ & $1.27^{+0.053}_{-0.10}$ & $0.35^{+0.067}_{-0.059}$ & $45.82^{+3.53}_{-4.1}$ & $1.11^{+0.05}_{-0.05}$ \\
$115^{+5}_{-5}$ & $31^{+1}_{-1}$ & $1.21^{+0.33}_{-0.50}$ & $0.34^{+0.050}_{-0.10}$ & $40.59^{+4.51}_{-4.15}$ & $1.08^{+0.05}_{-0.05}$ \\
$125^{+5}_{-5}$ & $24^{+1}_{-1}$ & $1.19^{+0.45}_{-0.67}$ & $0.21^{+0.042}_{-0.039}$ & $52.12^{+5.18}_{-3.62}$ & $0.63^{+0.05}_{-0.05}$ \\
 \hline
  \end{tabular}
    \end{center}
   \caption{Fitting results of the bursts 1-3 of observations 4U 1323-62. Here we provide the total unabsorbed flux from 0.001 - 100 keV in erg $cm^2$ $s^{-1}$.}
    \end{table*}

 \begin{table*}
   \begin{center}
   \begin{tabular} {c c c c c c c c}
 \hline 
& & LAXPC-10 &  & \\
\hline
& &  Burst-4 & &  \\
\hline
 Segments  & Rate (count/s) & $kT_{in}$ (keV)  & Flux $\times$ $10^{-9}$ & Radius (km)  & Chi-square \\
\hline

$5^{+5}_{-5}$ & $24^{+1}_{-1}$ & $1.83^{+0.08}_{-0.08}$ & $0.16^{+0.033}_{-0.033}$ & $45.20^{+0.82}_{-0.95}$ & $1.16^{+0.05}_{-0.05}$ \\
$15^{+5}_{-5}$ & $60^{+1}_{-1}$ & $2.15^{+0.18}_{-0.20}$ & $0.60^{+0.048}_{-0.049}$ & $87.56^{+1.19}_{-1.17}$ & $1.19^{+0.05}_{-0.05}$ \\
$25^{+5}_{-5}$ & $358^{+1}_{-1}$ & $2.27^{+0.05}_{-0.06}$ & $3.0^{+1.17}_{-0.069}$ & $96.10^{+0.32}_{-0.19}$ & $1.31^{+0.05}_{-0.05}$ \\
$35^{+5}_{-5}$ & $172^{+1}_{-1}$ & $2.08^{+0.08}_{-0.09}$ & $1.41^{+0.064}_{-0.065}$ & $94.45^{+0.35}_{-0.38}$ & $1.35^{+0.05}_{-0.05}$ \\
$45^{+5}_{-5}$ & $117^{+1}_{-1}$ & $1.80^{+0.09}_{-0.09}$ & $1.0^{+0.095}_{-0.2}$ & $90.55^{+0.45}_{-0.52}$ & $0.69^{+0.05}_{-0.05}$ \\
$55^{+5}_{-5}$ & $95^{+1}_{-1}$ & $1.91^{+0.13}_{-0.15}$ & $0.74^{+0.049}_{-0.049}$ & $90.29^{+1.15}_{-1.21}$ & $1.27^{+0.05}_{-0.05}$ \\
$65^{+5}_{-5}$ & $80^{+1}_{-1}$ & $1.72^{+0.15}_{-0.15}$ & $0.60^{+0.044}_{-0.044}$ & $88.48^{+1.09}_{-1.01}$ & $1.47^{+0.05}_{-0.05}$ \\
$75^{+5}_{-5}$ & $64^{+1}_{-1}$ & $1.63^{+0.16}_{-0.15}$ & $0.49^{+0.042}_{-0.042}$ & $86.11^{+0.79}_{-0.78}$ & $1.27^{+0.05}_{-0.05}$ \\
$85^{+5}_{-5}$ & $58^{+1}_{-1}$ & $1.58^{+0.15}_{-0.17}$ & $0.48^{+0.042}_{-0.040}$ & $86.05^{+0.75}_{-0.91}$ & $1.59^{+0.05}_{-0.05}$ \\
$95^{+5}_{-5}$ & $49^{+1}_{-1}$ & $2.18^{+0.30}_{-0.35}$ & $0.36^{+0.042}_{-0.043}$ & $79.12^{+0.52}_{-0.55}$ & $1.027^{+0.05}_{-0.05}$ \\
$105^{+5}_{-5}$ & $42^{+1}_{-1}$ & 1.20  & 0.32 & 77.07 & 2.74  \\
$115^{+5}_{-5}$ & $39^{+1}_{-1}$ & $1.83^{+0.39}_{-0.52}$ & $0.27^{+0.036}_{-0.037}$ & $71.21^{+0.32}_{-0.33}$ & $1.47^{+0.05}_{-0.05}$ \\
\hline
& & Burst-5  & &    \\
\hline
 Segments  & Rate (count/s) & $kT_{in}$ (keV)  & Flux $\times$ $10^{-9}$ & Radius (km)  & Chi-square \\
\hline

$5^{+5}_{-5}$ & $23^{+1}_{-1}$ & $0.85^{+0.052}_{-0.097}$ & $ 0.26_{-0.096}^{+0.051}$ & $39.83^{+10.12}_{-3.47}$ & $1.44^{+0.05}_{-0.05}$ \\
$15^{+5}_{-5}$ & $305^{+1}_{-1}$ & $2.25^{+0.069}_{-0.071}$ & $2.47_{-0.086}^{+0.087}$ & $94.79^{+1.66}_{-1.67}$ & $1.22^{+0.05}_{-0.05}$ \\
$25^{+5}_{-5}$ & $209^{+1}_{-1}$ & $2.15^{+0.082}_{-0.085}$ & $1.64_{-0.072}^{+0.073}$ & $99.00^{+2.26}_{-2.13}$ & $1.13^{+0.05}_{-0.05}$ \\
$35^{+5}_{-5}$ & $140^{+1}_{-1}$ & $2.034^{+0.11}_{-0.11}$ & $1.14_{-0.060}^{+0.060}$ & $90.12^{+2.47}_{-2.29}$ & $1.30^{+0.05}_{-0.05}$ \\
$45^{+5}_{-5}$ & $115^{+1}_{-1}$ & $1.95^{+0.13}_{-0.14}$ & $0.92_{-0.055}^{+0.074}$ & $87.86^{+2.74}_{-2.52}$ & $1.77^{+0.05}_{-0.05}$ \\
$55^{+5}_{-5}$ & $88^{+1}_{-1}$ & $1.65^{+0.13}_{-0.15}$ & $0.74_{-0.049}^{+0.049}$ & $85.42^{+3.02}_{-2.44}$ & $1.56^{+0.05}_{-0.05}$ \\
$65^{+5}_{-5}$ & $83^{+1}_{-1}$ & $1.79^{+0.16}_{-0.18}$ & $0.65_{-0.048}^{+0.048}$ & $83.11^{+3.23}_{-2.90}$ & $1.44^{+0.05}_{-0.05}$ \\
$75^{+5}_{-5}$ & $71^{+1}_{-1}$ & $2.00^{+0.18}_{-0.21}$ & $0.61_{-0.048}^{+0.049}$ & $81.52^{+3.46}_{-3.10}$ & $0.98^{+0.05}_{-0.05}$ \\
$85^{+5}_{-5}$ & $64^{+1}_{-1}$ & $1.92^{+0.27}_{-0.32}$ & $0.51_{-0.046}^{+0.047}$ & $77.13^{+3.73}_{-3.17}$ & $1.45^{+0.05}_{-0.05}$ \\
$95^{+5}_{-5}$ & $52^{+1}_{-1}$ & $1.95^{+0.27}_{-0.31}$ & $0.45_{-0.045}^{+0.046}$ & $74.17^{+4.02}_{-3.51}$ & $0.82^{+0.05}_{-0.05}$ \\
$105^{+5}_{-5}$ & $46^{+1}_{-1}$ & $1.35^{+0.34}_{-0.60}$ & $0.41_{-0.016}^{+0.041}$ & $56.92^{+6.96}_{-3.03}$ & $1.33^{+0.05}_{-0.05}$ \\
$115^{+5}_{-5}$ & $36^{+1}_{-1}$ & $2.48^{+0.74}_{-1.25}$ & $0.30_{-0.048}^{+0.19}$ & $58.44^{+5.21}_{-12.85}$ & $0.87^{+0.05}_{-0.05}$ \\
\hline
\hline
& & Burst-6  & &     \\
\hline
 Segments  & Rate (count/s) & $kT_{in}$ (keV)  & Flux $\times$ $10^{-9}$ & Radius (km)  & Chi-square \\
\hline
$5^{+5}_{-5}$ & $36^{+1}_{-1}$ & $2.51^{+0.17}_{-0.19}$ & $0.29_{-0.085}^{+0.10}$ & $64.34^{+1.04}_{-5.78}$ & $1.21^{+0.05}_{-0.05}$ \\
$15^{+5}_{-5}$ & $327^{+1}_{-1}$ & $2.28^{+0.069}_{-0.07}$ & $2.56_{-0.088}^{+0.088}$ & $95.46^{+1.72}_{-1.65}$ & $1.70^{+0.05}_{-0.05}$ \\
$25^{+5}_{-5}$ & $208^{+1}_{-1}$ & $2.05^{+0.075}_{-0.077}$ & $1.86_{-0.072}^{+0.072}$ & $94.36^{+2.1}_{-1.74}$ & $1.73^{+0.05}_{-0.05}$ \\
$35^{+5}_{-5}$ & $114^{+1}_{-1}$ & $1.94^{+0.11}_{-0.11}$ & $1.0_{-0.09}^{+0.07}$ & $88.55^{+4.33}_{-2.77}$ & $0.68^{+0.05}_{-0.05}$ \\
$45^{+5}_{-5}$ & $106^{+1}_{-1}$ & $1.88^{+0.12}_{-0.13}$ & $0.88_{-0.051}^{+0.051}$ & $88.43^{+3.26}_{-1.96}$ & $1.49^{+0.05}_{-0.05}$ \\
$55^{+5}_{-5}$ & $92^{+1}_{-1}$ & $1.86^{+0.16}_{-0.17}$ & $0.73_{-0.049}^{+0.049}$ & $86.94^{+3.07}_{0.51}$ & $1.42^{+0.05}_{-0.05}$ \\
$65^{+5}_{-5}$ & $82^{+1}_{-1}$ & $1.82^{+0.17}_{-0.19}$ & $0.61_{-0.045}^{+0.045}$ & $84.75^{+3.3}_{-2.9}$ & $1.12^{+0.05}_{-0.05}$ \\
$75^{+5}_{-5}$ & $61^{+1}_{-1}$ & $1.95^{+0.24}_{-0.26}$ & $0.48_{-0.042}^{+0.043}$ & $79.74^{+3.82}_{-0.36}$ & $1.46^{+0.05}_{-0.05}$ \\
$85^{+5}_{-5}$ & $54^{+1}_{-1}$ & $2.56^{+0.63}_{-0.60}$ & $0.42_{-0.056}^{+0.095}$ & $75.02^{+5.41}_{-7.07}$ & $1.71^{+0.05}_{-0.05}$ \\
$95^{+5}_{-5}$ & $47^{+1}_{-1}$ & $1.70^{+0.24}_{-0.28}$ & $0.38_{-0.038}^{+0.038}$ & $74.77^{+4.05}_{-3.55}$ & $1.07^{+0.05}_{-0.05}$ \\
\hline
    \end{tabular}
    \end{center}
   \caption{Fitting results of the bursts 4-6 of observations 4U 1323-62. Here we provide the total unabsorbed flux from 0.001 - 100 keV in erg $cm^2$ $s^{-1}$.}
    \end{table*}

The contribution from the persistent emission spectrum component is excluded from studying the variation of the neutron star radii and flux during the burst. Decoupling the burst component of the spectrum from the persistent part was difficult, as the various components were usually spectrally degenerated.

To generate energy spectra during the persistent emission pre-burst phase we selected 10 sec stretch of data prior to the thermonuclear burst. To fit the energy spectrum different model combinations (available in XSPEC) is used (i) tbabs(bbodyrad+powerlaw), (ii) tbabs(bbodyrad+diskbb), (iii) tbabs(bbodyrad+powerlaw+gaussion), (iv) tbabs(nthcomp+gaussion) and (v) tbabs(nthcomp). We observed that the model [tbabs*(nthcomp)] is best fit (lowest reduced $\chi^2$) to the persistent energy spectrum in 2-20 keV energy range. The hydrogen column density is fixed at $N_H$ = 4 $\times$ $10^{22}$ $cm^2$ \citep{2005boirin}.

The burst energy spectra in a 3-20 keV energy band were fitted with the standard approach: a blackbody model consisting of two parameters, a temperature ($kT_{bb}$) and normalization along with the persistent emission component of the corresponding burst. The parameters of the persistent emission component were frozen during the burst spectra. These spectral components took into account the nature of the seed photons (represented by the inp-type parameter that is equal to 0 for blackbody seed photons). This approach was used to understand the evolution of blackbody parameters.  The assumed blackbody emission well describes the burst spectra from the surface of the neutron star and comptonization emission from the extended ADC \citep{2001blchurch,2002chuuu}.

The unabsorbed bolometric flux (in the range 3-80 keV) was computed using the XSPEC function ``cflux". The total flux varies uniformly. The blackbody radius also varies consistently and show positively Pearson correlation with $kT\_{bb}$. $R_{bb}$ is the blackbody radius and $d_{10}$ is the source distance
in units of 10 kpc, $R_{bb}$ = {[N $\times$ $(distance)^2/4$$\pi$$flux$]$^{1/2}$} is obtained at the minimum distance of the source at 10 kpc.

The emission radii were estimated from the blackbody normalization. All the observed fit parameters have been shown in multi-panel plots with incremental time segments of each of the six bursts, as shown in figure \ref{multiple_fig8}. The first panel of each figure shows the variation in the count rate during the burst. The second panel shows the evolution of blackbody temperature. The third panel shows the total unabsorbed bolometric flux. The radius and reduced chi-squared shown in the fourth and fifth panel. The maximum temperature ($kT_{bb}$) observed in first X-ray burst is $3.05_{-0.32}^{+0.2}$ keV. However, we noticed the lower value of $kT_{bb}$ was $\sim$1.25 keV; all other segments in all bursts $kT_{bb}$ are varied in these two values. The 11th segments of B4 burst were not a good fit resulting in poor chi-square of 2.7.

\section{Discussion and conclusions}

We present the analysis of the AstroSat/LAXPC observation of the neutron star 4U 1323-62. LAXPC's payload detected six thermonuclear X-ray bursts and two dips from the source. Each burst had a similar wait time between consecutive bursts except the gap observed between the bursts B3-B4 and B5-B6, which is nearly double the wait time of consecutive bursts. We produced the linear and quadratic orbital ephemerids to find the value of the orbital period is $\sim$ 2.65 to 2.70 h, this was compatible with previous estimations, as shown in table \ref{instruments1}. 

The large effective area of the LAXPCs detector allowed us to generate the energy resolved burst profile at a higher energy band. We found that bursts were detected up to 20 keV. Above 20 keV, the LAXPC instruments showed a negligible count rate in the light curve of 4U 1323-62. However, the hard X-ray light curves during the bursts were found to be consistent with the pre-burst emission.  We observed strong energy dependence of burst decay times. The increase in the energy with decreasing the burst decay time, suggests the cooling of burning ashes \citep{2017galloway}.

The low frequency quasi-periodic oscillations (LFQPO) are observed in the PDS of the accreting NSs, non-pulsating NSs, and from the BH binaries. The phenomena of QPO frequency (0.6-2.4 Hz) for atoll sources depend on the photon energy. The observed QPO frequency  $\sim$ 1 Hz represents the high-inclination system  \citep{2015homanj}. The occurrence of QPOs is due to the physical and geometric change in the accretion disk. The disturbance of the inner accretion disk is the reason for the emission matter from the boundary layer or the surface of the neutron star.  

The state of the source was estimated by the normalization of components, which was highly variable during the bursts. During the soft state, the emission was from the accretion disc, and the harder one was from the boundary layer. Using our model, we investigated the constraints that can be obtained with the data of a much statistical precision than presently available. The bursts spectrum reasonably well described in the energy band 3.0 -20.0 keV by either a blackbody or power-law component. Time-resolved spectroscopy of the thermonuclear bursts indicated that the blackbody temperature varies from $\sim$3.0 to $\sim$1.0 as the burst's progress. The bolometric unabsorbed flux is found to decrease monotonically with the blackbody temperature during the thermonuclear bursts.  To averaging  all  the  values  of  flux,  the  mean  flux  for 4U  1323-62 was obtained i.e $\sim$ 1.80 $\times$ $10^{-9}$ erg $cm^2$ $s^{-1}$.

We also performed the time-resolved spectroscopy of burst using the standard technique. The evolution of temperature and radius indicates the Photosphere Radius Expansion (PRE). The rising phase of the bursts suggested that the emission during the burst reached the highest limit, while the persistent emission was at the lowest limit. A large increase in the burst flux during the rising phase was seen in very few neutron star LMXB systems such as GRS 1741.9-2853, MXB 1658-298. Our analysis showed that the radius expansion phase persists for 10-30 sec, which was slowly increasing as compared to the increase in radius in other LMXB systems like 4U 1608-52 and 4U 1728-34 
\citep{2016zhhang}.

The radius of the blackbody was highly consistent with the flux of each segment of the bursts. We extracted the apparent radius between $\sim$ 96 km - $\sim$ 40 km, under the assumption of an isotropically emitting spherical surface at a distance of 10 kpc. In our observations, it had been noticed that the flux continues to increase considering the maximum radii of the blackbody, such that total flux is strongly correlated with the radius.

\section{Acknowledgments}
We acknowledge the support from \textbf{Indian Space Research Organization (ISRO)} and 	\textit{AstroSat} science support cell in various aspect of instrument building, testing, software development and mission operation during payload verification phase. We thank LAXPC instrument team. JR wish to thank IUCAA for research facilities. 

\bibliographystyle{raa}
\bibliography{ref.bib}
\end{document}